\documentclass[prb,twocolumn,amsmath,amssymb]{revtex4-1}
\usepackage{graphicx}
\usepackage{sidecap}
\usepackage{bm}
\usepackage{amssymb}
\usepackage{amsmath}
\usepackage{dcolumn}
\usepackage{color}
\usepackage{epstopdf}
\usepackage{float}

\begin{document}

\title{\Large Zero-Field Ambient-Pressure Quantum Criticality in the Stoichiometric Non-Fermi Liquid System CeRhBi}

\author{\bf  \mbox{ V.\ K.\ Anand,$^{1,2,*}$ D.\ T.\ Adroja,$^{1,3,\dagger}$ A.\ D.\ Hillier,$^{1}$ K.\ Shigetoh,$^{4}$  T.\ Takabatake,$^{4}$} Je-Geun\ Park$^{5,6}$  K.\ A.\ McEwen,$^{7}$ J. H. Pixley$^{8,9}$ and Q. Si$^{10}$}

\affiliation{\mbox{$^1$ISIS Facility, Rutherford Appleton Laboratory, Chilton, Didcot, Oxon, OX11 0QX, United Kingdom}\\  \mbox{$^2$Helmholtz-Zentrum Berlin f\"{u}r Materialien und Energie GmbH, Hahn-Meitner Platz 1, D-14109 Berlin, Germany}\\ \mbox{$^3$Highly Correlated Matter Research Group, Physics Department, University of Johannesburg, P.O. Box 524,} Auckland Park 2006, South Africa\\  
\mbox{$^4$Department of Quantum Matter, ADSM, Hiroshima University, Higashi-Hiroshima, 739-8530, Japan}\\ \mbox{$^5$Department of Physics and Astronomy, Seoul National University, Seoul 08826, Korea}\\ \mbox{ $^6$Center for Correlated Electron Systems, Institute for Basic Science, Seoul 08826, Korea}\\  \mbox{ $^7$Department of Physics and Astronomy, University College London, Gower Street, London, WC1E 6BT, United Kingdom} \\
\mbox{$^8$Condensed Matter Theory Center and the Joint Quantum Institute, Department of Physics, University of Maryland,} College Park, Maryland 20742-4111, USA \\
\mbox{$^9$Center for Materials Theory, Department of Physics and Astronomy, Rutgers University, Piscataway, NJ 08854, USA}\\
\mbox{$^{10}$Department of Physics and Astronomy, Rice University, Houston, TX 77005, USA}\\  
\mbox{e-mail: $^{*}$vivekkranand@gmail.com, and $^{\dagger}$devashibhai.adroja@stfc.ac.uk}}

\date{\today}

\maketitle

{\bf The strange electronic state of a class of materials which violates the predictions of conventional Fermi-liquid theory of metals remains enigmatic. Proximity to a quantum critical point is a possible origin of this non-Fermi liquid (NFL) behavior, which is usually accomplished by tuning the ground state with non-thermal control parameters such as chemical composition, magnetic field or pressure. We present the spin dynamics study of a stoichiometric NFL system CeRhBi, using low-energy inelastic neutron scattering (INS) and muon spin relaxation ($\bm{\mu}$SR) measurements. It shows evidence for an energy-temperature ($\bm{ E/T}$) scaling in the INS dynamic response and a time-field ($\bm{t/H^\eta}$) scaling of the $\bm{\mu}$SR asymmetry function indicating a quantum critical behavior in this compound. The $\bm{ E/T}$ scaling reveals a local character of quantum criticality consistent with the power-law divergence of the magnetic susceptibility, logarithmic divergence of the magnetic heat capacity and $\bm{T}$-linear resistivity at low temperature. The NFL behavior and local criticality occur over a very wide dynamical range at zero field and ambient pressure without any tuning in this stoichiometric heavy fermion  compound is striking, making CeRhBi an exemplary model system amenable to in-depth studies for quantum criticality.}

{\bf Introduction:} Fermi-liquid (FL) theory, indispensable for our understanding of the low temperature properties of metals,  predicts the electrical resistivity $\rho \sim T^2$, heat capacity $C \sim T$ and magnetic susceptibility $\chi$  independent of temperature $T$. The essence of Fermi-liquid theory is the quasiparticle concept introduced by Landau \cite{Landau1956} which is an established paradigm of modern condensed matter physics.  Metals with diverse physical properties have been found to follow the FL theory, including even unusual ones like heavy fermion systems, with a very large renormalized effective quasiparticle mass. Nevertheless, there remains a class of materials whose strange electronic state cannot be understood within the FL picture, the high-Tc  cuprate superconductors being a well known example. Due to the strong coupling nature of the problem, the quasiparticle description breaks down \cite{Schofield1999, Stewart2001, Gegenwart2008, Si2010, Stockert2011}, the high-$T_{\rm c}$ cuprate superconductors being a well known example. These so-called non-Fermi liquid (NFL) materials exhibit characteristic low temperature properties such as $\rho \sim T^n$ ($1 \leq n < 2$),  $C/T \sim - \ln T$ or $C/T \sim a-b T^{1/2}$ and $\chi \sim -\ln T$ or $\chi \sim T^{-p}$ ($p <1$).  The presence of energy-temperature $E/T$ scaling in their dynamical spin susceptibility, with an anomalous exponent ($\alpha$), is another remarkable feature of some of the NFL systems that is characteristic of a strongly interacting quantum critical point \cite{Sachdev}.  Despite continuous efforts over the past two decades, the microscopic origin for the emergence of a NFL ground state is still not well understood. The single-ion multichannel Kondo effect, proximity to a $T=0$~K continuous quantum phase transition or a quantum critical point (QCP), and the Kondo disorder model are among the proposed theoretical models for NFL \cite{Stewart2001,Si2001, Coleman2001, Miranda2005, Lohneysen2007, Si2014}. The NFL behavior at a QCP is usually described by two theoretical scenarios: the conventional spin-density-wave (SDW) model of quantum criticality and the unconventional critical Kondo destruction model of local quantum criticality \cite{Si2001,Coleman2001, Lohneysen2007, Si2014}. A model based on the critical Fermi surface has also been proposed for NFL behavior at a QCP \cite{Senthil2008}. It is to be noted that $E/T$ scaling of the dynamical susceptibility holds only for  the unconventional critical Kondo destruction model of local quantum criticality as shown for CeCu$_{5.9}$Au$_{0.1}$  \cite{Schroder2000}, in contrast to the Gaussian fixed point of the $T=0$ SDW transition, where $E/T$ scaling is violated \cite{Hertz1976, Millis1993, Moriya1985}. Recently the role of geometrical frustration on a QCP has been investigated theoretically, which shows that magnetic frustration is a way to tune the quantum fluctuation in the heavy fermion materials and provides a global magnetic phase diagram  \cite {Si2006, JP2014}.

A QCP is usually accomplished by a non-thermal control parameter such as chemical composition, magnetic field or pressure that controls the amplitude of quantum fluctuations. Until recently most studies of NFL systems have been made on off-stoichiometric systems \cite{Stewart2001} in which the proximity to a QCP is accessed by tuning the chemical composition. However, the presence of inherent disorder in such systems makes it difficult to understand the individual roles of disorder and quantum criticality  in producing NFL. Therefore, investigations on clean stoichiometric NFL systems are very important for a better and comprehensive understanding of the origin of the NFL behavior and a direct comparison with the theoretical predictions. Field-tuned ${\rm YbRh_2Si_2 }$ is one such stoichiometric quantum critical NFL system that has been a prototypical system for intensive study, yielding results consistent with the critical Kondo destruction scenario \cite{Trovarelli2000, Gegenwart2002, Custers2003, Paschen2004, Friedemann2010}. The observation of NFL behavior in stoichiometric systems usually requires tuning. The stoichiometric heavy fermion superconductor $\beta$-${\rm YbAlB_4 }$ has recently been identified to show NFL features without any external tuning, the nature of quantum criticality in this material is still not clear  \cite{Nakatsuji2008, Matsumoto2011, Tomita2015, Pixley2012}. The experimental realization of NFL behavior in a stoichiometric material, that is consistent with local quantum criticality, both without magnetic field and at ambient pressure has been challenging. 

Here we present a stoichoimetric system, namely the Kondo lattice heavy fermion compound CeRhBi that exhibits non-Fermi liquid behavior \cite{Sasakawa2005} without any tuning and thus is an ideal system for in-depth study using a wide range of microscopic techinques without the constraints on account of tuning by field or pressure.  In order to understand the physics of NFL behavior in stoichiometric CeRhBi we have investigated the spin dynamics using inelastic neutron scattering (INS) and muon spin relaxation ($\mu$SR) techniques accessing a wide dynamic range. We find clear evidence of quantum critical scaling in the dynamic response of the INS and $\mu$SR asymmetry function, strongly indicating that the NFL behavior in this material originates from quantum criticality. Our results establish the stoichiometric ~~CeRhBi to be the first example of a Ce-based system in which a NFL behavior of local quantum critical origin is realized without any tuning. Further in the isostructural compound CePdSn,  the next-next-neighbour (NNN) exchanges are dominant and have opposite sign compared with NN exchange \cite {CePdSn} indicating the important role of the magnetic frustration. Such competing interactions are likely to be present in CeRhBi, suggesting that the local quantum criticality in CeRhBi is achieved through the magnetic frustration.

\begin{figure}
\includegraphics[width=\columnwidth]{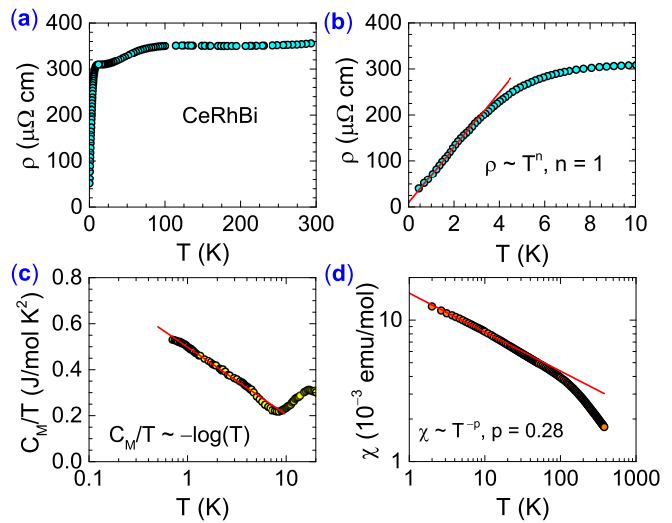}
\caption{\label{fig:NFL} {\bf Physical properties of CeRhBi: Evidence for non-Fermi liquid behavior:} (a) Electrical resistivity $\rho$ as a function of temperature $T$. (b) Low-$T$ $\rho(T)$ data showing the $T$-linear behavior of $\rho$. (c) Magnetic contribution of heat capacity $C_{M}(T)$ (after subtracting the lattice contribution using the heat capacity data of isostructural LaRhBi compound Ref.~[\onlinecite{LaRhBi}])  plotted as $C_{M}/T$ versus $T$ on a semi-log scale to show the logarithmic divergence of coefficient $\gamma$. (d) Magnetic susceptibility $\chi(T)$  measured in applied field $H = 1$~T plotted on a log-log scale to show the power law behavior, $\chi(T) \sim T^{-0.28}$. }
\end{figure}

{\bf Evidence for non-Fermi liquid behavior in ~~CeRhBi:} Clear evidence of non-Fermi liquid behavior is seen from the temperature dependence of the electrical resistivity, heat capacity and  magnetic susceptibility of CeRhBi shown in Fig.~\ref{fig:NFL}. While at high-temperature $\rho(T)$ exhibits a Kondo-lattice type feature [Fig.~\ref{fig:NFL}(a)], the low-$T$ $\rho(T)$ exhibits a $T$-linear behavior for $0.4 \leq T \leq 3$~K, i.e.\ $\rho(T) \sim T^{n}$ with $n = 1$ [Fig.~\ref{fig:NFL}(b)], a characteristic feature of non-Fermi liquid behavior. This NFL behavior is further supported by a logarithmic divergence in the magnetic heat capcity below 7~K,  $C_{M}(T)/T \sim -\ln(T)$ for $0.7 \leq T \leq 7$~K [Fig.~\ref{fig:NFL}(c)] with a very large electronic Sommerfeld coefficient $\gamma = C_{M}/T \approx 540$~mJ/mol\,K$^{2}$ at 0.7~K\@. Furthermore, the low-$T$ $ \chi(T)$ exhibits a power-law behavior below 20~K, $\chi(T) \sim T^{-p}$ with $p = 0.28$ [Fig.~\ref{fig:NFL}(d)]. Such a power-law divergence in magnetic susceptibility is predicted theoretically for a system near a locally critical quantum phase transition \cite{Si2001}. The broad peak in $\rho(T)$ near 100~K is due to the combined effect of Kondo effect and crystal field and the drop below 10~K is due to an onset of coherence. All these macroscopic properties reflect a local quantum critical behavior in CeRhBi, and motivated us to carry out inelastic neutron scattering measurements and examine the critical scaling of the spin dynamics.

\begin{SCfigure*}
\includegraphics[width=12.48cm]{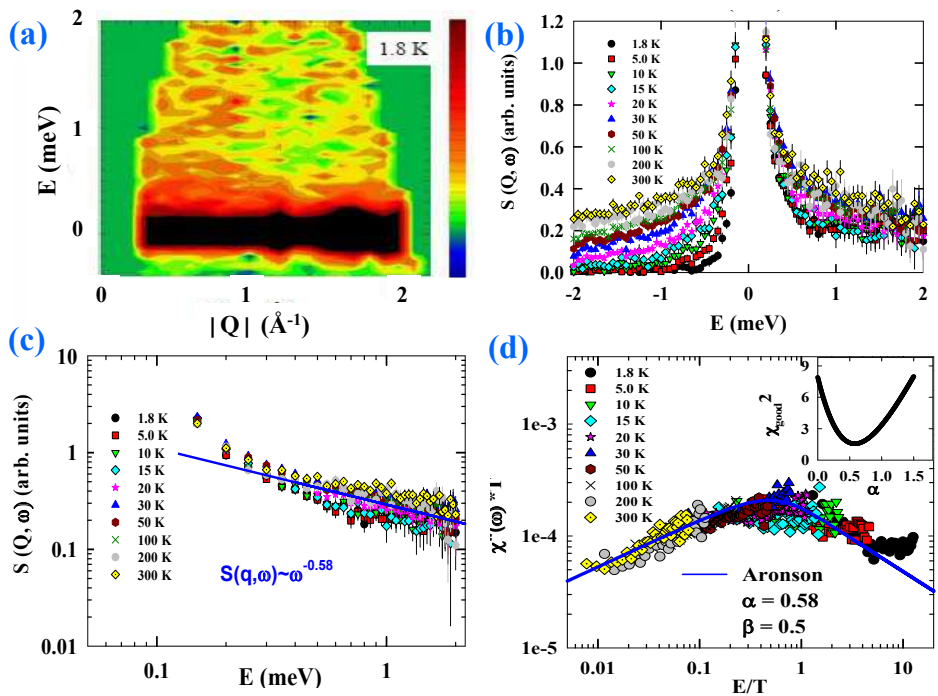}
\caption{\label{fig:INS1} {\bf Low energy inelastic neutron scattering from CeRhBi: Evidence for $\bm{ E/T}$ scaling: } (a) Inelastic neutron scattering response, a color-coded map of the intensity, energy transfer ($E$) versus momentum transfer ($Q$)  measured with the incident energy $E_i = 3.1$~meV at 1.8~K using the IN6 spectrometer at ILL. (b) $Q$-integrated ($0 < Q < 2$~{\AA}$^{-1}$) scattering intensity $S(Q,\omega)$ versus $E$ at different temperatures. (c) A log-log plot of $S(Q,\omega)$ data in (b) on the energy-loss part. The solid line shows $S(Q,\omega) \sim \omega^{-0.58}$ scaling behavior. (d) Dynamic susceptibility $\chi''(\omega, T)$ plotted as $\chi''(\omega, T) T^{\alpha}$ versus $E/T$ for $\alpha=0.58$. The solid curve is the fit according to Aronson scaling function in Eq.(\ref{eq:scaling}). The inset in (d) shows the plot of goodness of the scaling analysis ${\chi_{good}}^2$  {\it versus} ${\alpha}$ obtained from the independent scaling analysis theory \cite {KAWASHIMA1993}.}
\end{SCfigure*}

\noindent {\bf Low energy inelastic neutron scattering: $\bm{ E/T}$ scaling: } Figure~\ref{fig:INS1}(a) shows the color coded contour plot of the observed scattering from CeRhBi at $T=1.8$~K measured on the neutron spectrometer IN6 at ILL, Grenoble, with incident energy $E_i = 3.1$~meV, plotted as energy transfer ($E$) versus momentum transfer ($Q$), revealing a clear sign of quasi-elastic scattering. Figure~\ref{fig:INS1}(b) shows the $Q$-integrated ($0 < Q < 2$~{\AA}$^{-1}$) scattering intensity $S(Q,\omega, T)$ versus $E$ for CeRhBi for  $1.8 \leq T \leq 300$~K summed over all scattering angles between 10$^{\circ}$ and 135$^{\circ}$. A narrow quasi-elastic linewidth is seen at 1.8~K\@. The magnetic scattering is almost $Q$-independent for all the measurements between 1.8 and 300~K\@. 

According to the fluctuation-dissipation theorem, the measured scattering intensity $S(Q,\omega,T)$ is related to the imaginary part of the dynamic susceptibility $\chi''(Q, \omega)$,  \cite{Xu2013}
\begin{equation}
S(Q,\omega, T) =  \frac{\hbar }{\pi g^2 \mu_{\rm B}^2}\, \frac{\chi''(Q,\omega, T)}{1-\exp(-\hbar \omega /k_{\rm B} T)}
\label{eq:INS_S}
\end{equation}
with $\chi''(Q,\omega, T) = \omega F^2(Q) \chi'(0,0, T)  P(Q,\omega,T)$ according to the Kramers-Kronig relation, where $F(Q)$ is the magnetic form factor, $\chi'(0,0, T) = \chi_0(T)$ is the static bulk susceptibility and $ P(Q,\omega,T) $ is the normalized spectral function.

A double logarithmic plot of the magnetic scattering $S(Q,\omega, T)$ as a function of $E$ is shown in Fig.~\ref{fig:INS1}(c) for all the measured temperatures. The $S(Q,\omega, T)$ between 0.25 and 2~meV at all temperatures are linear in $E$ on a log-log scale suggesting a power law behavior. The magnetic scattering thus follows a scaling behavior, $S(Q,\omega,T) \sim \omega^{-\alpha}$ with $\alpha = 0.58$ ($\omega$ is related to $E$ by $E = \hbar \omega)$. A very similar scaling behavior has also been observed in other non Fermi liquid systems such as UCu$_{5-x}$Pd$_x$ \cite{Aronson1995}, CeRh$_{0.8}$Pd$_{0.2}$Sb \cite{Park2002a}, Ce$_{0.7}$Th$_{0.3}$RhSb \cite{So2002} and CePd$_{0.15}$Rh$_{0.85}$ \cite{Adroja2007} with the exponent $\alpha = 0.33, 0.77$, 0.33 and 0.6, respectively. 

Further, we find a clear evidence of $E/T$ scaling in the imaginary part of the dynamic susceptibility. The $\chi''(\omega, T)$ obtained is shown in Fig.~\ref{fig:INS1}(d) as $\chi''(\omega, T) T^{\alpha}$ versus $E/T$ plot. The $\chi''(\omega, T)$ data between 0.2 and 2 meV at $1.8 \leq T \leq 300$~K collapse onto a single curve. This confirms the universal $E/T$ scaling behavior of $\chi''(\omega, T)$. The $\chi''(\omega, T)$ data are well described by the scaling relation $\chi''(\omega, T) T^{\alpha} \sim  f(\omega/T)$ with $\alpha = 0.58$. The solid curve in Fig.~\ref{fig:INS1}(d) represents the scaling function proposed by Aronson {\it et al}\ \cite{Aronson1995}.
\begin{equation}
\chi''(\omega, T) T^{\alpha} = (T/\omega)^{\alpha} \tanh(\omega/\beta T)
\label{eq:scaling}
\end{equation}
for  $\alpha = 0.58$ and $\beta = 0.5$. The $E/T$ scaling behavior has been observed in several NFL systems, however with different values of $\alpha$, and sometimes with different choice of scaling function $ f(\omega/T)$. In quantum spin glasses UCu$_{5-x}$Pd$_{x}$ \cite{Aronson1995} and Sc$_{1-x}$U$_{x}$Pd$_3$ \cite{Wilson2005} $\alpha = 1/3$ and 0.2, respectively. In antiferromagnetic QCP systems CeCu$_{6-x}$Au$_{x}$ \cite{Schroeder1998}, CeRh$_{0.8}$Pd$_{0.2}$Sb \cite{Park2002a}, Ce$_{0.7}$Th$_{0.3}$RhSb \cite{So2002} and Ce$_2$PdIn$_8$ \cite{Tran2012} $\alpha = 0.75$, 0.77, 0.33 and 1.5, respectively. In the ferromagnetic QCP system CeRh$_{0.85}$Pd$_{0.15}$ \cite{Adroja2007} $\alpha = 0.6$. The reason for the different values of the exponent in different compounds is not well understood. The wide variation in the $\alpha$ value might have a relation with the different origin to the quantum criticality, distance from the QCP in the phase space and the dimensionality including the amount of chemical disorder. 

In order to further confirm the value of $\alpha$ from our analysis, we used a model free analysis method \cite {KAWASHIMA1993} that allows to estimate the goodness of the scaling ${\chi_{good}}^2$  {\it versus} ${\alpha}$ from the INS data in Fig. 2(b). The plot of ${\chi_{good}}^2$  {\it versus} ${\alpha}$ shown in the inset of Fig.2(d) gives ${\alpha}$=0.58(8) for a minimum value of  ${\chi_{good}}^2$ validating our $E/T$ scaling analysis. The value of our exponent estimated from $E/T$ scaling (0.58) is close the theoretical value 0.72 predicted by a model based on a local QCP in an anisotropic Kondo lattice \cite {Grempel}. The scaling exponent should be the same in the dc-susceptibility and that obtained from $E/T$ scaling of the neutron data. We attribute the observed difference due to an anisotropic magnetic susceptibility, which could give preferred orientation in the dc-susceptibility data. The single crystal susceptibility of NFL CeRh$_{2}$Ga shows two different values of the scaling exponents in the single crystal susceptibility, $\chi_{\parallel}\sim$ T$^{-0.48}$ (for H${\parallel}$c-axis) and $\chi_{\perp}\sim$ T$^{-0.79}$ 
(for H${\perp}$c-axis) \cite {Chen}. In this respect the single crystal susceptibility data of CeRhBi would be enlightening.

The INS data measured with high incident energy neutrons show evidence for crystal field excitations (see Supplementary material)  within the $J=5/2$ ground multiplet of Ce$^{3+}$ ion which indicates that with increasing energy the $E/T$ scaling breaks down and the system becomes localized.

\begin{figure*}
\includegraphics[width=\textwidth]{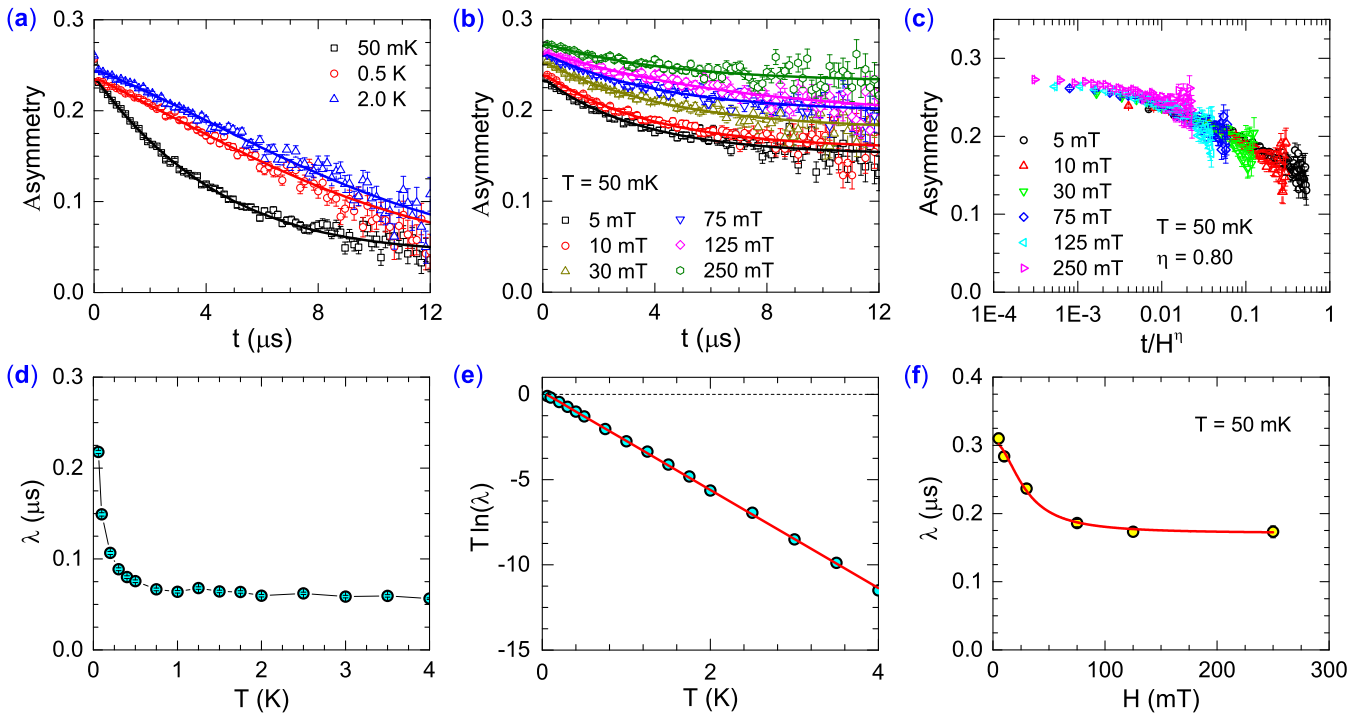}
\caption{\label{fig:MuSR1} {\bf Muon spin relaxation in CeRhBi: Evidence for time-field scaling:} (a) The zero field $\mu$SR spectra  collected at the indicated temperatures. The solid curves are the fits to the data by  Eq.~(\ref{eq:ZF_MuSR}). (b) The longitudinal field $\mu$SR spectra of CeRhBi collected at indicated fields at 50~mK\@. The solid curves are the fits to the data by  Eq.~(\ref{eq:LF_MuSR}). (c) Time-field scaling of longitudinal field $\mu$SR data in (b), $G_z(t)$ versus $t/H^\eta$  for $\eta=0.80$. (d) Temperature $T$ dependence of the depolarization rate $\lambda$ obtained from the zero field $\mu$SR data. (e) $T\ln(\lambda)$ versus $T$ plot. The solid line is the fit according to Arrhenius-like activated behavior. (f) Field $H$ dependence of $\lambda$ obtained from the longitudinal field $\mu$SR data. The solid curve is the fit according to Eq.~(\ref{eq:Lambda_tau}).}
\end{figure*}

\noindent {\bf Muon spin relaxation: time-field scaling:} The representative $\mu$SR spectra collected at various temperatures between 50~mK and 4~K while warming the sample in zero field (ZF) are shown in Fig.~\ref{fig:MuSR1}(a). The absence of frequncy oscillations or any loss in initial asymmetry indicates no long range order down to 50~mK, though the muon depolarization rate seems to increase as the temperature is lowered indicating that the muons sense the presence of quantum fluctuations. The ZF $\mu$SR spectra are well described by a Gaussian Kubo-Toyabe function multiplied by an exponential decay,
\begin{equation}
 G_z(t)= A_{0}\left[ \frac{1}{3}+\frac{2}{3}(1-\sigma^2 t^2)\, e^{-\sigma^2 t^2/2}  \right] e^{-\lambda t} + A_{\rm BG}
\label{eq:ZF_MuSR}
\end{equation}
where A$_{0}$ is the initial asymmetry from the sample and $A_{\rm BG}$ is the asymmetry contribution from the silver sample holder, $\sigma$ is the relaxation rate of the Gaussian channel of depolarization which provides a measure of the distribution of static magnetic fields at the muon site arising from the nuclear moments that are static on the time scale of the muon lifetime, and $\lambda$ is the relaxation rate of the Lorentzian channel accounting for the dynamic magnetic fields due to fluctuating electronic moments. The fits of the ZF $\mu$SR spectra by Eq.~(\ref{eq:ZF_MuSR}) are shown by the solid curves in Fig.~\ref{fig:MuSR1}(a). The fit parameter $\sigma$ was found to be constant within the error bar therefore for final fits we fixed it to $\sigma = 0.0735$~$\mu$s$^{-1}$. $A_0$ was also found to remain fairly $T$-independent with a value of about 0.20, however a sharp increase is seen in $\lambda(T)$ as $T$ decreases [Fig.~\ref{fig:MuSR1}(d)]. The $\lambda(T)$ diverges as $T \rightarrow 0$~K, and shows an activated behavior $\lambda(T) = \lambda_0 \exp(-E_g/k_{\rm B}T)$ or equivalently $T \ln(\lambda) = T \ln(\lambda_0) - E_g/k_{\rm B}$, where $E_g$ is the energy gap and $k_{\rm B}$ is the Boltzmann's constant. A linear fit of $T\ln(\lambda)$ versus $T$ plot [Fig.~\ref{fig:MuSR1}(e)] yields $E_g = 0.14(3)$~K\@. Thus the spin dynamics seems to be thermally activated, indicating the presence of low energy spin fluctuations in CeRhBi. A very similar behavior was also observed for CeInPt$_4$ with activation energy of 2.9~mK \cite{Hillier2007}.

The application of a longitudinal magnetic field is found to decouple the nuclear component such that the LF $\mu$SR spectra are well described by a simple Lorentzian function,
\begin{equation}
 G_z(t)= A_{0} e^{-\lambda t} + A_{\rm BG}.
\label{eq:LF_MuSR}
\end{equation}
The LF $\mu$SR spectra collected at 50~mK for $5 \leq H \leq 250$~mT and fits to them by Eq.~(\ref{eq:LF_MuSR}) (solid curves) are shown in Fig.~\ref{fig:MuSR1}(b). The $H$ dependence of the electronic relaxation $\lambda$ is shown in Fig.~\ref{fig:MuSR1}(f). The $\lambda$ initially decreases rapidly with increasing $H$ and eventually tends to be constant. The $\lambda(H)$ allows us to estimate the spin auto correlation time $\tau_c$ according to the fluctuation-dissipation theorem which relates $\tau_c(H)$ to the imaginary component of the local $Q$-independent $f$-electron dynamic susceptibility $\chi^{\prime\prime}(\omega)$, which for $\hbar \omega \ll k_{\rm B}T$ is given by $ \tau_c (H) = ({k_{\rm B}T}/{\mu_{\rm B}^2}) {\chi^{\prime\prime}(\omega)}/{\omega}$. The $\lambda(H)$ data follow \cite{Amato1997}
\begin{equation}
 \lambda(H) = \lambda_0+ \frac{2\gamma_\mu^2 \langle H_{\rm loc}^2 \rangle \tau_c}{1+\gamma_\mu^2 H^2 \tau_c^2}
\label{eq:Lambda_tau}
\end{equation}
where $\lambda_0$ is the $H$-independent depolarization rate, $\langle H_{\rm loc}^2 \rangle $ is the time average of the second moment of the time-varying local field $H_{\rm loc}(t)$ at muon sites due to the fluctuations of neighboring Ce 4f moments. The fit of the $\lambda(H)$ by Eq.~(\ref{eq:Lambda_tau})  obtained for $\lambda_0=0.17(1)~\mu$s$^{-1}$, $\surd {\langle H_{\rm loc}^2 \rangle} = 1.5(1)$~mT, and $\tau_c = 4.2(6)\times 10^{-8}$~s is shown  by the solid red curve in Fig.~\ref{fig:MuSR1}(f). The value of $\tau_c$ reveals a slow spin dynamics which is caused by the quantum critical fluctuations.

Furthermore we find that the LF $\mu$SR data follow characteristic time-field scaling  $G_z(t, H)= G_z(t/H^\eta)$, where the exponent $\eta$ provides information about spin-spin dynamical autocorrelation, $\eta < 1$ for power-law correlation and $\eta > 1$ for stretched exponential correlation \cite{Keren1996,MacLaughlin2001,MacLaughlin2004}. The $G_z(t)$ versus $t/H^\eta$ plot of $H$-dependent LF $\mu$SR data at $T =50$~mK  are shown in Fig.~\ref{fig:MuSR1}(c). The time-field scaling of $\mu$SR data at $T=50$~mK is evident from Fig.~\ref{fig:MuSR1}(c) for $\eta=0.80(5)$. The value of $\eta=0.80(5) <1$ suggests a power-law correlation and hence cooperative and critical spin fluctuations in CeRhBi. This kind of time-field scaling in LF $\mu$SR has been observed in a few other NFL systems \cite{MacLaughlin2004}, such as in UCu$_{5-x}$Pd$_{x}$ with $\eta=0.35$ for $x=1$ and $\eta=0.7$ for $x=1.5$ \cite{MacLaughlin2001}. The stoichiometric antiferromagnetic QCP system ${\rm YbRh_2Si_2 }$ \cite{Ishida2003} shows time-field scaling with $\eta \approx 1$. The ferromagnetic QCP system CePd$_{0.15}$Rh$_{0.85}$ \cite{Adroja2008} shows scaling for  $\eta=1.0(1)$. 

\noindent {\bf Discussion and summary:} 
The linear $T$ dependence of $\rho$, logarithmic divergence of $C(T)/T$ and power-law divergence of $\chi(T)$ provide clear evidence of a NFL ground state in CeRhBi. Further, the INS data reveal universal $E/T$ scaling of $\chi''(\omega, T)$ and $\mu$SR spectra show the time-field scaling. While the time-field scaling indicates slowing of spin-dynamics due to quantum critical fluctuations, the $E/T$ scaling and power-law divergence of $\chi(T)$ provide strong evidence for local quantum criticality \cite{Si2001} in CeRhBi. Importantly, the quantum critical behavior is realized at zero field and ambient pressure. As there is no obvious inherent disorder in CeRhBi (being stoichiometric and crystallographically well ordered) the Kondo disorder model is of not much relevance here. Moreover, the multichannel Kondo effect model too is not relevant as the $\chi(T)$ does not follow the expected logarithmic divergence. One natural route to this continuous transition at ambient conditions could be provided by the underlying magnetic frustration in CeRhBi, which has tunned the microscopic exchange constants to place it at this QCP. 

The theory of local quantum criticality\cite{Si2001,Coleman2001} was formulated to explain the origin of NFL behavior and $E/T$ scaling in CeCu$_{6-x}$Au$_{x}$ \cite{Schroeder1998,Lohneysen1994,Schroeder2000}. The local quantum criticality is associated with the breakdown of the Kondo screening, resulting in a collapse of the Fermi surface \cite{Si2014}, however the off-stoichiometric nature of CeCu$_{6-x}$Au$_{x}$ limits a precise study of the Fermi surface in this material. Evidence for the critical destruction of Kondo screening (Fermi-surface collapse) has been found from Hall-effect measurements \cite{Paschen2004,Friedemann2010} on the stoichiometric heavy fermion ${\rm YbRh_2Si_2 }$ that exhibits a magnetic field-tuned quantum criticality and NFL ground state \cite{Trovarelli2000,Gegenwart2002,Custers2003}. A similar critical Kondo destruction has also been observed in field-tuned quantum critical heavy fermion system ${\rm Ce_3Pd_{20}Si_6}$ \cite{Custers2012}. Similar evidence has come from quantum-oscillation measurements of the Fermi surface in the heavy fermion ${\rm CeRhIn_5 }$ \cite{Jiao2015}.

To summarize, our spin dynamics study of stoichiometric non-Fermi-liquid heavy fermion CeRhBi provides clear evidence for universal quantum critical scaling in this compound over a very wide dynamic range. CeRhBi thus emerges as a uniquely significant Kondo lattice system, situated at a  quantum critical point at zero field and ambient pressure in stoichiometric form. The observed full value of the effective paramagnetic moment and very small Kondo temperature (T$_{K} \sim$9 K ~Ref.~[\onlinecite{LaRhBi}]) with the absence of magnetic ordering down to 50 mK, indicate that the magnetic frustration is more important than Kondo effect leading to in the observed QCP and NFL behavior of CeRhBi.
That the unconventional local quantum criticality in CeRhBi is realized without any tuning makes it a promising system for comprehensive studies to unravel the puzzle of quantum criticality and NFL physics. Of particular interest will be to probe the Fermi surface using techniques such as angle-resolved photoemission spectroscopy (ARPES) and to examine the criticality of the Fermi surface. From a more general materials perspective, the critical Kondo destruction, which is the most favored underlying mechanism for the dynamical $E/T$ scaling, has so far been typically observed in field-tuned QCP and the observation of Kondo destruction at a field-free and ambient-pressure QCP has been lurking. Our results establish CeRhBi as a much-needed model compound for the experimental study of such a critical Kondo destruction at a QCP under ambient conditions. 

\vspace{0.5cm}

\noindent {\bf Methods} 

\noindent Approximately 15~g polycrystalline samples of CeRhBi and its nonmagnetic reference compound LaRhSb were synthesized by melting the high purity elements in a sealed tungsten crucible and subsequent annealing at 900~$^\circ$C for a week.  The magnetic measurements were performed using a commercial superconducting quantum interference device magnetometer (MPMS, Quantum Design Inc.). The heat capacity and electrical resistivity were measured using a physical properties measurement system (PPMS, Quantum Design Inc.). The low energy inelastic neutron scattering measurements were carried out on the IN6 time-of-flight spectrometer using an incident energy $E_i = 3.1$~meV at ILL, Grenoble, France. The energy resolution at the elastic position was 0.08 meV. The high energy inelastic neutron scattering measurements were carried out on the HET time-of-flight spectrometer using $E_i = 60$~meV at the ISIS pulsed neutron and muon facility of the Rutherford Appleton Laboratory, Didcot, U.K\@. The $\mu$SR measurements were carried out  in both longitudinal and transverse geometry on the MuSR spectrometer at ISIS.  For the $\mu$SR measurement the powdered sample of CeRhBi was mounted on a high purity silver holder using GE varnish.

\vspace{0.5cm}

\noindent  {\bf Acknowledgements} 

\noindent We thank A. P. Murani for his assistance in the neutron scattering measurements on IN6. We thank A.M.\ Strydom and B.D. Rainford for helpful discussions. V.K.A., D.T.A.\ and A.D.H.\ acknowledge financial assistance from CMPC-STFC grant number CMPC-09108. J.G.P.\ was supported by the research programme of Institute for Basic Science (IBS-R009-G1). J.H.P.\  acknowledge funding from LPS-MPO-CMTC. The work at Rice University was supported by the ARO grant No. W911NF-14-1-0525 and the Robert A. Welch Foundation grant No. C-1411, with travel support provided by the 
 NSF grant No. DMR -1611392.

\vspace{0.5cm}

\noindent  {\bf Author contributions} 

\noindent  D.T.A.\ and  V.K.A.\ designed the research project. K.S.\ and T.T.\ synthesized and measured the magnetic susceptibility, heat capacity and electrical resistivity. D.T.A.\ and K.A.M.\ carried out the inelastic neutron scattering (INS) measurements, discussed the results with J.-G.P. \ and D.T.A.\ analyzed the INS data. D.T.A.\ and A.D.H.\ carried the muon spin relaxation measurements. V.K.A., D.T.A.\ and A.D.H.\ analyzed the muon data.  J.H.P.\ and Q.S.\ provided the theoretical inputs to understand the experimental data.  V.K.A.\ and D.T.A.\ drafted the manuscript and prepared the supplementary materials. All authors actively discussed the results and interpretations, and participated in the writing and review of the manuscript.

\vspace{0.5cm}

\noindent  {\bf Additional information}

\noindent Supplementary information is available.\\
\noindent Correspondence and requests for materials should be addressed to D.T.A.\ or V.K.A. 

\vspace{0.5cm}

\noindent  {\bf Competing financial interests}

\noindent The authors declare no competing financial interests.\\

\clearpage

\section*{\huge Supplementary Material} 

\section{Crystal Structure}

\begin{figure}[b]
\includegraphics[width=\columnwidth]{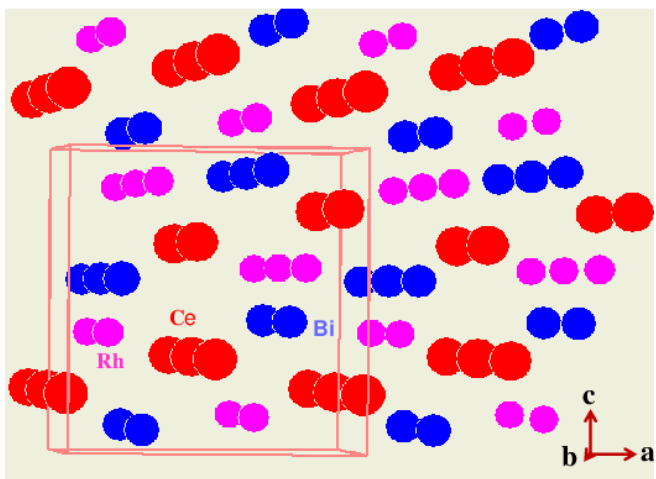}
\caption{\label{fig:Stru} {\bf Crystal structure of CeRhBi:} The orthorhombic TiNiSi-type crystal structure of CeRhBi.  The big red spheres are Ce atoms,  small purple spheres are Rh atoms and small blue spheres are Bi atoms.}
\end{figure}

Figure~\ref{fig:Stru} shows the crystal structure of CeRhBi, which crystallizes in the orthorhombic TiNiSi-type structure with space group $Pnma$ (No. 62). The orthorhombic lattice parameters are  7.431, 4.727, and 7.991 $\AA$. The shorter distances for Ce-Ce are 3.832, 3.8796 $\AA$, Ce-Rh are 3.1271, 3.2984 $\AA$ and Ce-Bi are 3.1417, 3.3601 $\AA$. The Ce-atoms forms a zig-zag chain along the b-axis. The crystal structure of CeRhBi is same as that of Kondo insulators CeRhSb and CeNiSn and both of them are low-carrier-density Kondo insulators with enhanced spin fuctuations but no magnetic ordering down to 100 mK. Each cerium atom is surrounded by a polyhedron of twenty atoms, i.e. 8 Ce, 6 Rh and 6 Bi atoms. The coordination numbers of Rh and Bi atoms are 10 and their coordination polyhedra are trigonal prisms. Generally, a distinguishing feature for the TiNiSi type compounds is the different coordination of of the metal and nonmetal atoms. 

\section{Magnetic Susceptibility and Magnetization}

\begin{figure}
\includegraphics[width=\columnwidth]{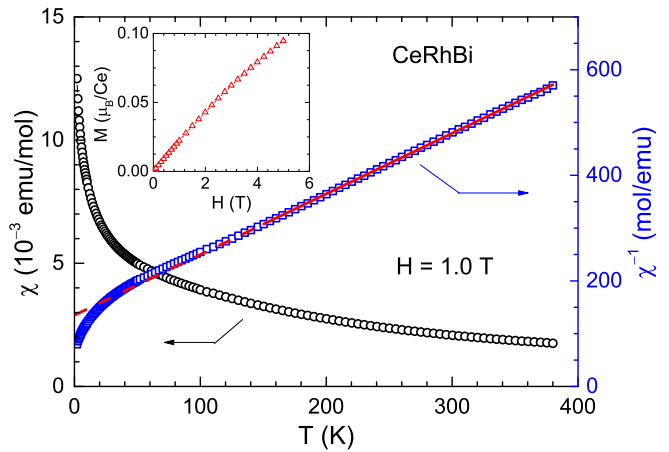}
\caption{\label{fig:Chi} {\bf Magnetic susceptibility and isothermal magnetization of CeRhBi:} Magnetic susceptibility $\chi$ and its inverse $\chi^{-1}$ as a function of temperature $T$ for in $2~{\rm K} \leq T \leq 380$~K measured at an applied field $H = 1.0$~T\@.  The solid red line represents the fit to Curie-Weiss behavior in and its extrapolation is shown by dashed line. The inset shows the isothermal magnetization $M$ as a function of $H$.}
\end{figure}

Figure~\ref{fig:Chi} shows the temperature $T$ dependence of magnetic susceptibility $\chi$ of CeRhBi measured at an applied field $H = 1.0$~T\@. At high temperature the $\chi(T)$ data follow Curie-Weiss behavior $\chi(T) = C/(T-\theta_{\rm p})$. A linear fit of the inverse susceptibility $\chi^{-1}(T)$ in \mbox{$150 \leq T \leq 380$~K} yields the paramagnetic Weiss temperature $\theta_{p} = -117(1)$~K and Curie constant $C= 0.870(2)$~emu/mol\,K coresponding to an effective magnetic moment $\mu_{\rm eff} = 2.64\,\mu_{B}$. The $\mu_{\rm eff}$  which is close to the theoretical value of $2.54\,\mu_{B}$ expected for a free Ce$^{3+}$ ion, indicates a Ce$^{3+}$ state of Ce ions at high temperatures. The isothermal magnetization $M(H)$ data at 2~K are shown in the inset of Fig.~\ref{fig:Chi}. It is seen that the $M$ attains only about $0.1\,\mu_{B}$/Ce at 5.0~T which is much smaller than the expected saturation moment of $2.14\,\mu_{B}$/Ce for Ce$^{3+}$ ion. 
The slight nonlinearity in $M(H)$ could be the result of the crystal electric field effect and/or Kondo coherence.

\section{Heat Capacity}

\begin{figure}
\includegraphics[width=\columnwidth]{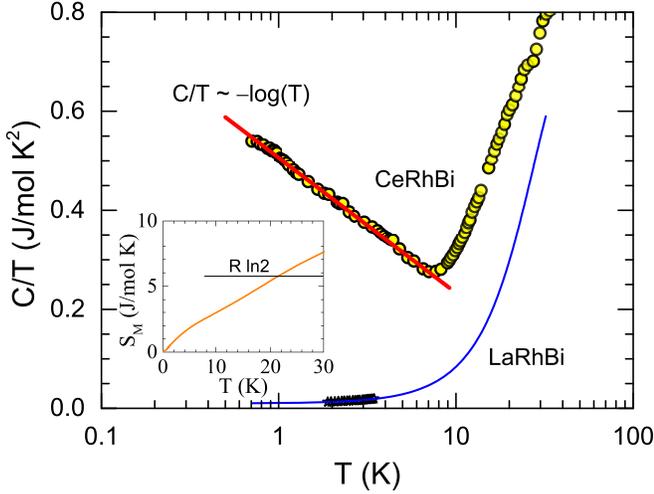}
\caption{\label{fig:HC} {\bf Heat capacity of CeRhBi:} Heat capacity $C$ of CeRhBi plotted as $C/T$ versus temperature $T$ on a semi-log scale to show the logarithmic divergence of coeffcient $\gamma$. The $C(T)$ of LaRhBi \cite{Yoshii1997} is also shown. The solid blue curve is the calculated $C(T)$ for nonmagnetic reference LaRhBi according to Debye model of lattice heat capacity. The inset shows the magnetic entropy $S_{\rm M}(T)$.}
\end{figure}

Figure~\ref{fig:HC} shows the heat capacity $C(T)$ of CeRhBi plotted as $C/T$ versus $T$ CeRhBi. The low-$T$ $C(T)$ data show a clear logarithmic divergence of Sommerfeld coefficient $\gamma(T) = C(T)/T \sim -\ln(T) $ at $T\leq 7.0$~K. The magnetic contribution to heat capacity $C_{\rm M}$ was obtained by subtracting off the phonon contribution using the reported low temperature heat capacity of LaRhBi \cite{Yoshii1997}. The low-$T$ normal state $C(T)$ of LaRhBi is well described by $C(T) = \gamma T + \beta T^3$ with $\gamma = 10$~mJ/mol\,K$^2$ and  $\beta = 0.74$~mJ/mol\,K$^4$. A Debye temperature of $ \Theta_{\rm{D}} =199$~K is obtained from the value of $\beta$. In order to obtain the phonon contribution over an extended region of temperature we calculated Debye lattice heat capacity (blue curve in Fig.~\ref{fig:HC}) according to \cite{Gopal1966} 
\begin{equation}
C_{\rm{V\,Debye}}(T) = 9 n R \left( \frac{T}{\Theta_{\rm{D}}} \right)^3 {\int_0^{\Theta_{\rm{D}}/T} \frac{x^4 e^x}{(e^x-1)^2}\,dx},
\label{eq:Debye_HC}
\end{equation}
for $ \Theta_{\rm{D}} =199$~K, which was then subtracted off from $C(T)$ of CeRhBi. The $C_{\rm M}(T)$ obtained this way is shown in Fig.~1(c) of manuscript. The magnetic entropy $S_{\rm M}(T)$ obtained from the $C_{\rm M}(T)$ data is shown in the inset of Fig.~\ref{fig:HC}. From the magnetic entropy one can estimate Kondo temperature which is twice the temperature at which $S_{\rm M} = 0.4 \,R \ln2$, accordingly we obtain $T_{\rm K } \approx 14$~K\@.

\section{Crystal Field States}

\begin{figure}
\includegraphics[width=3in]{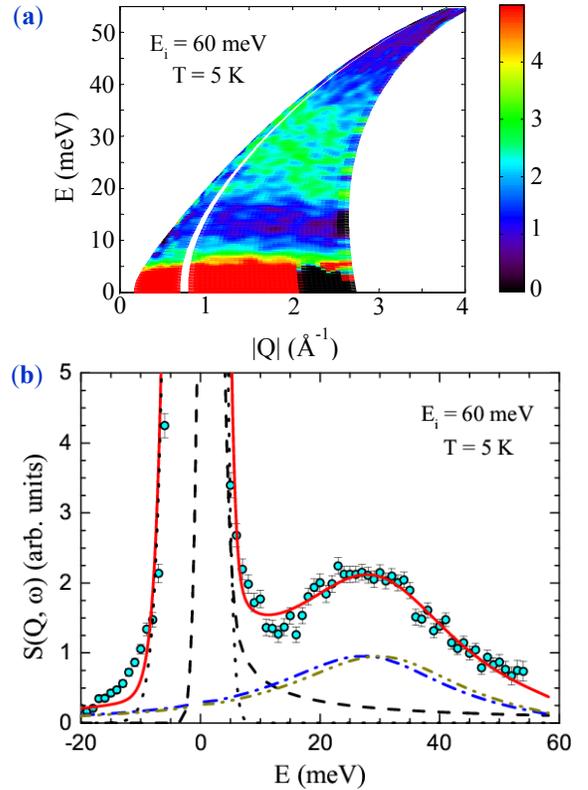}
\caption{\label{fig:CEF} {\bf Inelastic neutron scattering: Crystal field states of CeRhBi:} (a) Magnetic inelastic neutron scattering response, a color-coded map of the intensity, energy transfer ($E$) versus momentum transfer ($Q$) of CeRhBi measured with the incident energy $E_i = 60$~meV at 5~K using the HET spectrometer at ISIS. (b) The magnetic scattering intensity $S(Q,\omega)$ versus $E$ at 5~K. Solid red curve is the fit based on the crystal field model. The dashed and dash-dotted lines represent the different components of fit. }
\end{figure}

Figure~\ref{fig:CEF}(a) shows the color coded contour plot of the observed inelastic neutron scattering from CeRhBi measured on HET with the incident energy $E_i = 60$~meV at 5~K, plotted as momentum transfer ($Q$) versus energy transfer ($E$). The phonon scattering was subtracted off using the data of nonmagnetic reference LaRhSb measured with the identical condition. The presence of a sharp broad inelastic excitation near 30~meV is evident from the contour plot. This shows that with increasing energies the $E/T$ scaling breaks down and the system becomes localized. The inelastic excitations near 30~meV can be attributed to crystal electric field (CEF) excitation of the $J=5/2$ ground multiplet of Ce$^{3+}$ ion, which splits into three doublets under the influence of the low point symmetry crystal field potential. The $Q$-dependent energy integrated intensity between 25 and 35 meV follows the Ce$^{3+}$ magnetic form factor squared [$F^2(Q)$] indicating that at high incident energies the inelastic excitations in CeRhBi result essentially from the single-ion CEF transitions.

The magnetic scattering intensity $S(Q,\omega)$ versus $E$ at 5~K obtained from HET data is shown in Fig.~\ref{fig:CEF}(b). We analyze these high energy INS data by a model based on crystal field. The crystal field is expected to remove the six-fold degeneracy of ground state of Ce$^{3+}$ ($^2F_{5/2}$) spliting it into three doublets. Therefore one would expect two CEF excitations from the two excited states, however $S(Q,\omega)$ in Fig.~\ref{fig:CEF}(b) shows only a rather broad CEF excitation (near 30 meV) which could be due to two unresolved excitations because of a very small splitting energy between the two excitated states. CeRhBi forms in TiNiSi-type orthorhombic structure (space group $Pnma$, No.~62).  In this orthorhombic structure CeRhBi has a triclinic point symmetry which can be approximated to a trigonal $D_{3d}$ symmetry such that the CEF Hamiltonian can be described as
\begin{equation}
H_{\rm{CEF}}=  B_2^0O_2^0+B_4^0O_4^0+B_4^3C_4^3
\end{equation}
where $O_n^m$ are the Stevens' equivalent operators and $B_n^m$ are phenomenological CEF parameters. The red solid curve in Fig.~\ref{fig:CEF}(b) represents the fit to this CEF model. Also shown are the different components of the fit. The CEF analysis of INS data yield the phenomenological parameters (in meV) $B_{2}^{0}$ = -0.0156, $B_{4}^{0 }$= 0.0479, and  $B_{4}^{3}$ = -1.648 and splitting of the three doublets to be 0, 28.68 and 31.38~meV.  The ground state wave function is a mixture of $|\pm 1/2 \rangle$ and $|\pm 5/2 \rangle$.

\end{document}